\documentclass[doublecol]{epl2}
\usepackage{amssymb}

\title{Influence of initial distributions on robust cooperation in evolutionary Prisoner's Dilemma}
\shorttitle{Influence of initial distributions etc.} %Insert here a short version of the title if it exceeds 70 characters

\author{X.-J. Chen\inst{1,2} \and F. Fu\inst{1,2} \and L. Wang\inst{1,2}\thanks{E-mail: \email{longwang@pku.edu.cn}}}
\shortauthor{X.-J. Chen \etal}  %Insert here the first author if authors exceed 70 characters

\institute{
  \inst{1} Intelligent Control Laboratory, Center for Systems and Control, Department of Mechanics and Space Technologies, College of Engineering, Peking University, Beijing 100871,
  China\\
  \inst{2} Department of Industrial Engineering and Management, College of Engineering, Peking University, Beijing 100871, China
}

\abstract{We study the evolutionary Prisoner's Dilemma game on
scale-free networks for different initial distributions. We
consider three types of initial distributions for cooperators and
defectors: initially random distribution with different
frequencies of defectors; intentional organization with defectors
initially occupying the most connected nodes with different
fractions of defectors; intentional assignment for cooperators
occupying the most connected nodes with different proportions of
defectors at the beginning. It is shown that initial
configurations for cooperators and defectors can influence the
stationary level of cooperation and the evolution speed of
cooperation. Organizations with the vertices with highest
connectivity representing individuals cooperators could exhibit
the most robust cooperation and drive evolutionary process to
converge fastest to the high steady cooperation in the three
situations of initial distributions. Otherwise, we determine the
critical initial frequencies of defectors above which the
extinction of cooperators occurs for the respective initial
distributions, and find that the presence of network loops and
clusters for cooperators can favor the emergence of cooperation.}

\begin{document}

\maketitle

\section{Introduction}

Evolutionary game theory has become an important tool for
investigating cooperative behavior of biological, ecological,
social and economic systems \cite{1,2}. The Prisoner' Dilemma game
(PDG) is one of the most commonly employed games for this purpose.
Originally, in the PDG, two individuals adopt one of the two
available strategies, cooperate or defect; both receive $R$ under
mutual cooperation and $P$ under mutual defection, while a
cooperator receives $S$ when confronted to a defector, which in
turn receives $T$, where $T>R>P>S$ and $T+S<2R$. Under these
conditions it is best to defect for rational individuals in a
single round of the PDG, regardless of the opponent strategy.
However, mutual cooperation would be preferable for both of
individuals. Thus, the dilemma is caused by the selfishness of the
individuals.

However, the unstable cooperative behavior is opposite to the
observations in the real world. This disagreement thus motivates
to find under what conditions the cooperation can emerge on the
PDG. Graph theory provides a natural and very convenient framework
to study the evolution of cooperation in structured populations.
In well-mixed populations, each individual interacts with each
other individual. The average payoff of defectors is greater than
the average payoff of cooperators and the frequency of cooperators
asymptotically vanishes. In other structured populations, each
individual occupies one vertex and individuals only interact with
their neighbors in a social network. Several studies have reported
the cooperation level on different types of networks
\cite{3,4,5,6,7}. Nowak and May introduced a spatial evolutionary
PDG model in which individuals located on a lattice play with
their neighbors, and found that the spatial effect promotes
substantially the emergence of cooperation \cite{3}. Santos
\textit{et al}. have studied the PDG and Snowdrift game (SG) on
scale-free networks and found that comparing with the regular
networks, scale-free networks provide a unifying framework for the
emergence of cooperation \cite{6}. Notably, scale-free networks
where the degree distribution follows a power law form are highly
heterogeneous, and the heterogeneity of the network structure can
promote cooperation. However, the puzzle of cooperation on social
networks is unanswered yet. Recently, the roots of the diverse
behavior observed on scale-free networks are explored \cite{8,9}.
Cooperators can prevail by forming network clusters, where they
help each other on heterogeneous networks \cite{10}. In scale-free
networks, the majority of nodes have only a few links, while a
small number of nodes with high connectivity (hubs) are well
connected to each other. This extremely inhomogeneous connectivity
distribution results in the robustness of scale-free networks
\cite{11}. As a result, the presence of hubs and relative
abundance of small loops for cooperators in scale-free networks
can promote the level of cooperation.

From these results on scale-free networks, it seems that
cooperation can be affected by the initial distribution for
cooperators (C) and defectors (D), such as randomly or
intentionally distributions, individuals initially assigned with
equal or unequal probability to be C or D. Similarly, a special
initial distribution for C and D may exhibit a robust cooperation
on scale-free networks. However, in most literature, initial
strategies of individuals are randomly assigned with the same
probability to be C or D. Here, we remove the setting and are
interested in investigating the evolution of cooperation for
different initial distributions on scale-free networks. The paper
is organized as follows. In the next section, we describe the
evolutionary game model as well as networks in detail. And then
simulation results and analysis are provided in the third section.
Finally, conclusions are given in the fourth section.

% \appendix

\section{The model}

Firstly, we construct scale-free networks using the Barab\'{a}si
and Albert model (BA) which is considered to be the typical model
of the heterogeneous networks \cite{12}. Starting from $m_0$
vertices which are connected to each other, at each time step one
adds a new vertex with $m$ ($m\leq m_0$) edges that link the new
vertex to $m$ different vertices already present in the system.
When choosing the vertices to which the new vertex connects, one
assumes that the probability $P_i$ that a new vertex will be
connected to vertex $i$ depends on the degree $k_i$ of vertex $i$:
$P_i=k_i/\sum_jk_j$. After $t$ time steps this algorithm produces
a grape with $N=t+m_0$ vertices and $mt$ edges. Here, we set
$m=m_0=2$ and network size $N=3000$ for all the simulations. Thus,
the average degree of this network model can be given
$\bar{k}=2m=4$.

After constructing networks, each site of the network is occupied
by an individual. Each individual who is a pure strategist can
only follow two simple strategies: cooperate and defect. In one
generation, each individual plays a PDG with its neighbors
simultaneously, and collects payoffs dependent on the payoff
matrix parameters. The total payoff of a certain individual is the
sum over all interactions in one generation. Following common
practice \cite{3,13}, we use a simplified version of PDG, make
$T=b$, $R=1$ and $P=S=0$, where $b$ represents the advantage of
defectors over cooperators, being typically constrained to the
interval $1<b<2$. Let us represent the individuals' strategies
with two-component vector, taking the value $s=(1, 0)^T$ for
C-strategist and $s=(0, 1)^T$ for D-strategist. Therefore, the
total payoff $P_x$ of a certain individual $x$ can be written as
\begin{equation}
P_x=\sum_{y\in \Omega_x} s_x^T As_y, \label{eq1}
\end{equation}
where the sum runs over all the neighboring sites of $x$, and
$\Omega_{x}$ is the set of neighbors of element $x$.
%The total payoff of each individual can be calculated by
%Eq.~(\ref{eq1}).

During the evolutionary process, each individual is allowed to
learn from one of its neighbors and update its strategy in each
generation. Following previous works \cite{13,14}, each individual
chooses one individual randomly from its neighbors. After choosing
a neighbor $y$, the individual $x$ adopts the selected $y$
neighbor's strategy in the next generation with a probability
depending on their total payoff difference as
\begin{equation}
W_{s_x\leftarrow s_y}=\frac{1}{1+\exp{[(P_x -P_y)/K]}},\label{eq2}
\end{equation}
where $K$ characterizes the noise effects, including fluctuations
in payoffs, errors in decision, individual trials, etc. And $P_x$,
$P_y$ denote the total payoffs of individuals $x$ and $y$,
respectively. Here, $K$ is set to $0.125$ for the total payoffs.
Furthermore, the results remain unaffected with different values
of the parameter $K$.

\section{Simulations and discussion}

In the following, we will show the simulation results carried out
for a population of $N=3000$ individuals occupying the vertices of
the scale-free networks with $\bar{k}=4$. The above model is
simulated with synchronous updating. Eventually, the system
reaches a dynamic equilibrium state. The equilibrium frequencies
of C are obtained by averaging over the last $1000$ generations
after a transient time of $10000$ generations. In what follows,
three situations of initial distributions for C and D will be
considered: (1) defectors are randomly distributed to occupy the
network vertices; (2) defectors on purpose occupy the highly
connected nodes; (3) defectors are intentionally assigned to
occupy the nodes with small connectivity. In these respective
situations, the effects of different initial frequencies of
defectors $f_{ID}$ on the emergence of cooperation are
subsequently investigated, too. In situations (2) [situation (3)],
nodes in the scale-free networks are sorted by decreasing
(increasing) number of links that each node contains. There are
instances where groups of nodes contain identical numbers of
links. Where this occurs, they are arbitrarily assigned a position
within that groups. For example, the node rank $r$ denotes the
position of a node on this ordered list and $1\leq r\leq N$
\cite{15}. Initially, when $r$ defectors occupy the highly
connected nodes, they just occupy the $r$ nodes with highest
connectivity in the networks ; while $r$ defectors occupy the
nodes with small connectivity, they just occupy the $r$ nodes with
smallest connectivity in the networks, and thus $f_{ID}=r/N$ is
the initial frequency of D. The evolution of the frequency of C as
a function of $b$ and $f_{ID}$ for different initial distributions
has been computed. To this end, each data point results from an
average over $30$ realizations of both the networks and same
initial distributions.
\begin{figure*}
\centering
\includegraphics[width=15cm]{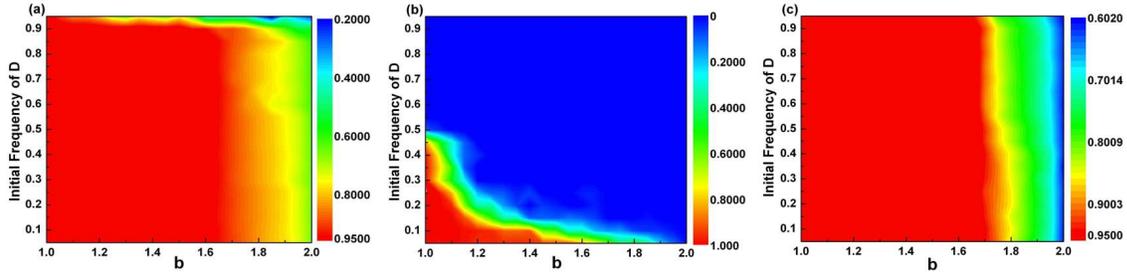}
\caption{(Color Online) Evolution of cooperation in scale-free
network with $\bar{k}=4$. Results for the fraction of C at
equilibrium in the population are plotted as a  contour, drawn as
a function of two parameters: $b$ and $f_{ID}$. (a) random
distributions with different initial frequencies of D; (b)
different initial fractions of D which occupy nodes with high
connectivity; (c) different initial percentages of D which occupy
nodes with small connectivity.}
\label{fig.1}
\end{figure*}
\begin{figure*}
\centering
\includegraphics[width=15cm]{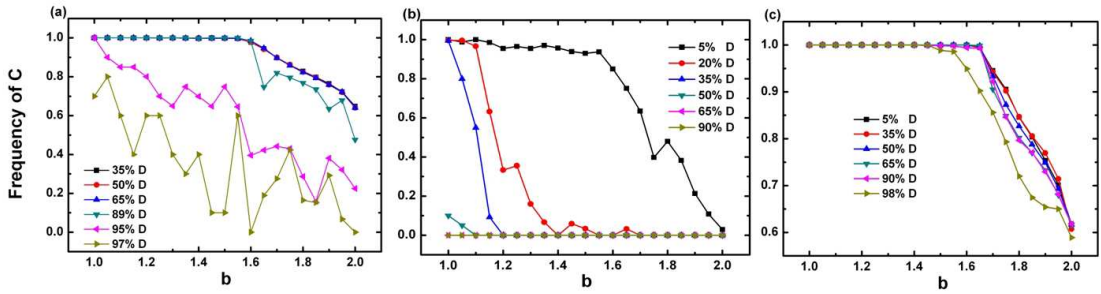} \caption{(Color
Online) Frequency of C at equilibrium as a function of the
parameter $b$ for different distributions with different values of
$f_{ID}$.} \label{fig.2}
\end{figure*}
\begin{figure}
\centering
\includegraphics[width=7.5cm]{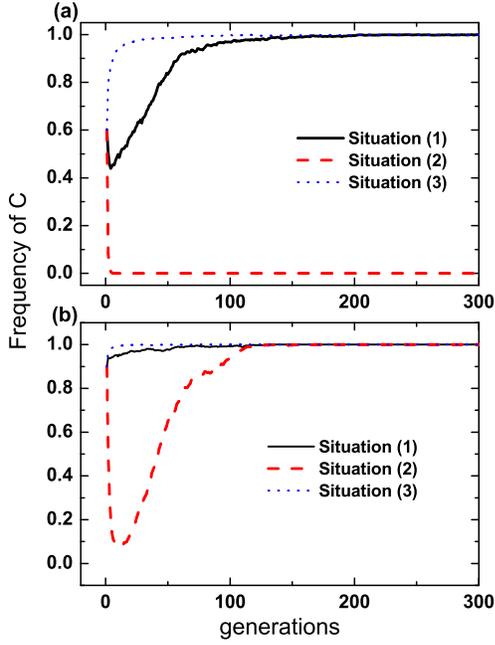}
\caption{(Color Online) Frequency of C at equilibrium as a
function of evolution generations for different values of $b$ and
initial distributions with the same $f_{ID}$. (a) $b=1.4$ and
$60\%$ cooperators at the beginning; (b) $b=1.1$ and $90\%$
cooperators at the beginning.}
\label{fig.3}
\end{figure}
\begin{figure}
\centering
\includegraphics[width=7.5cm]{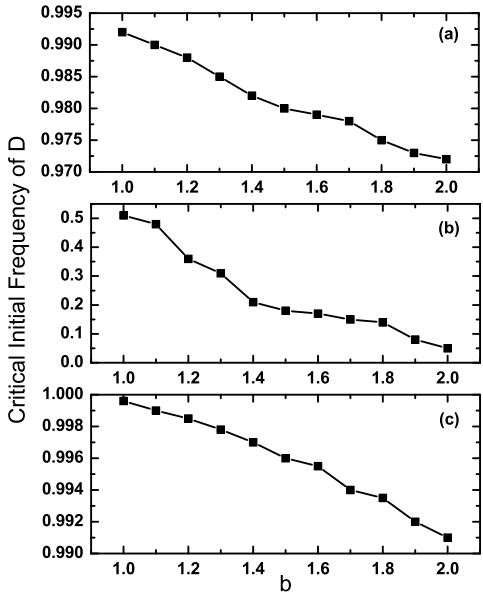}
\caption{Critical frequency of D for cooperators to vanish in the
PDG as a function of the parameter $b$ for different situations.}
\label{fig.4}
\end{figure}

Fig.~\ref{fig.1} shows the simulation results in the PDG for
different initial distributions as a contour plot. Clearly, in
fig.~\ref{fig.1}(a) we have found that the cooperation level
becomes poorer when the initial frequency of D increases for a
given fixed $b$. Especially, the cooperation level begins to
fluctuate and decreases intensively when initial frequency of D is
large and near one for high values of $b$. While cooperators
dominate over the most ranges of $b$ and $\rho_{ID}$ in this
situation. In fig.~\ref{fig.1}(b), cooperation strongly depends on
the values of $f_{ID}$, and defectors dominate over the most
ranges of $b$ and $f_{ID}$. In fig.~\ref{fig.1}(c), a certain
amount of cooperation can emerge and remain stable even for high
initial frequency of D, and cooperators prevail over the most
ranges of $b$ and $f_{ID}$. A comparison of the different results
for different initial frequencies of D is shown in
fig.~\ref{fig.2}. We depict the cooperation level as a function of
the parameter $b$. In fig.~\ref{fig.2}(a), we have found that the
equilibrium frequency of C begins to fluctuate and decrease for
high values of $b$ when the initial frequency of D is high. And
the cooperation level remains stable for small values of $b$.
Additionally, when $f_{ID}$ increases and approaches one,
cooperation fluctuates intensively and cooperators dies out
finally. As shown in fig.~\ref{fig.2}(b), cooperation is strongly
inhibited as $b$ increases when defectors are not wiped out. There
are larger oscillations and cooperation is sensitive to initial
frequency of D when defectors occupy the nodes with highest
connectivity at the beginning. Moreover, cooperators vanish when
the initial frequency of D is more than $50\%$ no matter what the
value of $b$ is. Fig.~\ref{fig.2}(c) exhibits a robust and
favorable cooperation for different initial frequencies of D. Even
if a small number of cooperators initially occupy the rich nodes,
it still leads to a high cooperation level. The frequency of C
decreases slowly for a high initial frequency of D over the whole
region of $b$. The cooperative behavior is robust against
defector's invasion in this situation. From fig.~\ref{fig.2}, we
know that different initial frequencies of D and distributions can
result in different levels of cooperation. In addition, in
comparison with the two other situations, the situation that
cooperators occupy the rich nodes, presents much more robust
cooperation in this respect that high cooperation remains for
almost any temptation. It is shown that the time evolution of
cooperation in PDG for different values of $b$ and initial
distributions with the same $f_{ID}$ in fig.~\ref{fig.3}. It is
found that situation (3) that cooperators occupy the most
connected nodes at the beginning makes evolutionary process
converge much fastest to the equilibrium state of $100\%$
cooperators in the three situations, while situation (2) that
defectors firstly occupy the nodes with highest connectivity
provides much harsher condition for the emergence of cooperation
than the two other situations and makes cooperation level drop
much fast. Situation (3) promotes the emergence of cooperation and
can speed up the evolution of cooperation. Fig.~\ref{fig.4} shows
the critical frequency of D for cooperators to vanish in the PDG
as a function of $b$ for the three types of initial distributions
and also illuminates these results. When the initial frequency of
D is higher than the critical frequency of D, cooperators vanishes
or decreases intensively to extinction. For an arbitrary value of
$b$, the critical frequency of D in situation (3) is always higher
than those in situation (1) and (2). Initial ratios of C in one
certain distribution for C and D can affect the cooperation level;
otherwise, initial distributions for C and D also influence the
emergence of cooperation and the evolution speed of cooperation.
\begin{figure}
\centering
\includegraphics[width=7.5cm]{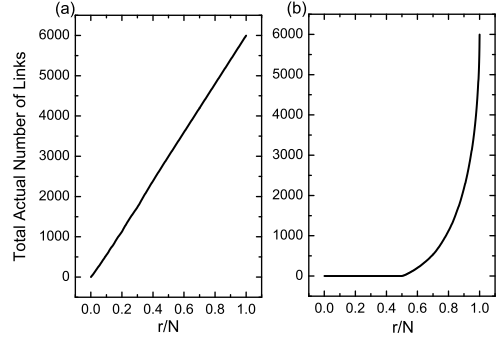}
\caption{The total actual number of links among $r$ nodes against
$r/N$ with $N=3000$ and $m=m_0=2$ in scale-free networks, where
$r$ represents the node rank. (a) $r$ nodes with the highest
connectivity in the networks; (b) $r$ nodes with the smallest
connectivity in the networks. Each data point of the curves
results from 10 different network realizations.}
\label{fig.5}
\end{figure}

These simulation results can be understood in the following way.
In scale-free networks, there are a large number of nodes which
have only a few links, and there are small number of links among
these less connected nodes; while there are a small number of
nodes with large numbers of links, these most connected nodes or
hubs are generally very well connected to each other (see
fig.~\ref{fig.5}). The connectivity between these hubs in the
networks can be crucial for the emergence of cooperation for the
PDG \cite{9,10,11,14,16}. Based on these results, some
corresponding explanations on our results can be provided. At
first, we discuss the random initial distribution with different
fractions of D. When the initial percentage of D is small, nodes
with high connectivity will be occupied by defectors with much
smaller probability. In this case, individuals using strategy C
representing highly connected nodes communicate with each other
and form loop and main cluster structures, and hence the high
levels of cooperation can emerge. Therefore, the probability, with
which most connected vertices are occupied by cooperators,
decreases when the initial fraction of D increases. Then clusters
of cooperators may be cut off (fragmented) from the main compact
cluster, but there are still some loops and fragments for
cooperators. In this state there is a systematic drop of
cooperation at the beginning, nevertheless it tends to rise again
in the long run, thereby, cooperation falls but can remain at a
high level. While the initial percentage of D is more than the
critical frequency, it is still possible for a small number of C
players to occupy the nodes with high connectivity although the
probability is so small, since strategies C and D are randomly
distributed among all the players. Thus, there are large
oscillations when the initial frequency of D approaches one,
because it is increasingly difficult for the cooperators occupying
most connected nodes to communicate with each other in this state.
And then we investigate the situation that defectors occupy the
nodes with highest connectivity at the beginning. In other words,
cooperators initially occupy vertices having only a few links. In
fig.~\ref{fig.5}(b), it shows that there are few actual links
among about half of the nodes which are almost the least connected
nodes. When the initial frequency of D is more than $50\%$ in
situation (2), it is not possible to form network clusters for
cooperators where cooperators can help each other, and defectors
are grouped in several clusters, then cooperators lose more and
more elements from their outer layer along with the increment of
evolution generations, therefore, cooperators can not survive no
matter what the value of $b$ is. Nevertheless, only small isolated
pieces can be formed for cooperators when the fraction of D is
less than the critical frequency, since defectors occupy the most
connected nodes. Thus, it results in that cooperation falls
intensively and can not remain stable. However, a high level of
cooperation is sustainable just for small values of $b$, because
in this case defectors have not much advantage over cooperators.
For $b\thicksim1$, cooperators are equivalent to defectors, then
the level of cooperation is not strongly susceptible to the
initial distribution for C and D. In fact, in all generations
cooperation falls rapidly at the beginning, then cooperators
sometimes recover but not always for small values of $b$ in
situation (2). For large $b$, cooperation always falls and never
recovers. Therefore, in this state cooperation drops rapidly and
it needs much time to revert cooperation if cooperation can
recover finally. Accordingly, cooperators vanish at the stationary
state over the most regions of $b$ and $f_{ID}$.  The situation,
that individuals C intentionally assigned to represent the
vertices with high connectivity at the beginning, is analyzed
finally. In this case, it is easy for cooperators to form giant
compact network clusters and loops. Even if a small number of
cooperators occupy the most connected nodes, there are a large
number of loops and some tiny compact clusters for cooperators;
conversely, defectors are not organized in these clusters, where
cooperators can help each other and defectors can not invade. The
presence of clusters and loops in the connectivity structure for
cooperators sustains the high level of cooperation even for a high
value of $b$, and in all generations they can favor cooperation at
the beginning, and drive evolutionary process to converge fast to
the high stationary cooperation level. Therefore, cooperators
dominate over the entire ranges of $b$ and $f_{ID}$, and the
cooperative behavior is robust against defectors' invasion in this
situation.

Scale-free networks have most of their connectivity clustered and
looped in a few nodes, therefore, initial assignments for C and D
can affect the cooperative behavior and the evolution speed of
cooperation. The configuration that cooperators initially occupy
the most connected nodes, presents the much more robust
cooperation than the two other ones and can speed up the evolution
of cooperation, in comparison with the two other different initial
configurations. Moreover, cooperators can prevail by forming
network clusters and loops, where they can assist each other.
These results are independent of the size of the populations $N$.

\section{Conclusions}

In summary, we have studied the cooperative behavior of the
evolutionary PDG on scale-free networks for different initial
distributions, and also found that the presence of network loops
and clusters for cooperators can favor cooperation. Cooperators
dominate over the most range of $b$ with different initial
frequencies of D when strategies C and D are randomly distributed
among the populations; a poor and unstable cooperation level can
be established at equilibrium in the state that the vertices with
high connectivity represent defectors at the beginning; while a
very robust and favorable cooperation can be exhibited in the
situation that the highly connected nodes are occupied by
cooperators at the beginning. The situation that cooperators
initially occupy the most connected nodes provides the most robust
cooperation in the three situations of initial distributions for C
and D. Additionally, it is found that the configuration that
cooperators occupy the most connected nodes at the beginning can
speed up the evolution of cooperation; while the situation that
defectors occupy the most connected nodes drives cooperation to
drop fast and difficultly recover. And the critical frequencies of
D for cooperators to vanish corresponding to initial distributions
have been respectively determined. Some qualified explanations
based on the property of scale-free networks are given for these
phenomenon. Therefore, our results shows that initial
configurations for C and D with different ratios of D at the
beginning can affect the levels of cooperation on heterogeneous
networks. Moreover, our work may be helpful in exploring the roots
of the emergence of cooperation on heterogeneous networks.

\acknowledgments This work was supported by National Natural
Science Foundation of China (NSFC) under grant No. 60674050 and
No. 60528007, National 973 Program (Grant No. 2002CB312200),
National 863 Program (Grant No. 2006AA04Z258) and 11-5 project
(Grant No. A2120061303).

\end{document}